\documentclass[aps,pra,twocolumn,groupedaddress,showpacs,superscriptaddress,amssymb,amsmath, color]{revtex4-2}
\usepackage{graphicx}
\usepackage{amsmath}
\usepackage{hyperref}
\usepackage{xcolor}
\hypersetup{
    colorlinks = true,
    linkcolor =blue,
	citecolor=green, 
	urlcolor=blue 
}
\newcommand{\pa}[1]{{\mathrm{p}_{#1}}}

\newcommand{\co}{\mathrm{c}}
\newcommand{\x}{\mathrm{x}}
\newcommand{\y}{\mathrm{y}}
\newcommand{\po}{\mathrm{xy}}
\newcommand{\Lo}{\mathrm{L}}
\newcommand{\ex}{\mathrm{ex}
}
\begin{document}

\title{Quantum error correction via multi-particle discrete-time quantum walk}
%
\author{Ryo Asaka}
\email{asaka@rs.tus.ac.jp}
\affiliation{Department of Physics, Tokyo University of Science, Kagurazaka 1--3, Shinjuku-ku, Tokyo 162--8601, Japan}

\author{Ryusei Minamikawa}
\email{ryusei.minamikawa@aist.go.jp}
\affiliation{National Metrology Institute of Japan (NMIJ), National Institute of Advanced Industrial Science and Technology (AIST), Tsukuba, Ibaraki 305-8563, Japan}

\begin{abstract}
    We propose a scheme of quantum error correction that employs a multi-particle quantum walk defined on nested squares, each hosting a single particle.
    In this model, each particle moves within its own distinct square through iterations of three discrete-time operations:
    (i) $\mathcal{C}$: each particle updates its two-level internal {\it coin} state,
    (ii) $\mathcal{S}$: it either shifts to an adjacent vertex or stays put, depending on the coin state,
    (iii) $\mathcal{N}$: it interacts with another particle if these particles arrive at the nearest-neighbor vertices of the two adjacent squares, acquiring a phase factor of $-1$.
    Because a single particle represents a three-qubit state through its position and coin state, Shor's nine-qubit code is implemented using only three particles, with two additional particles for syndrome measurement.
    Notably, our proposal would lead to ultrafast and resource-efficient quantum error correction by taking the continuous limit of the discrete-time iterations of $\mathcal{C}\rightarrow\mathcal{S}\rightarrow\mathcal{N}\rightarrow\mathcal{C}\rightarrow\cdots$.
    Note that the scheme is also resilient against a unified correctable noise model presented in the companion paper~(arXiv:2604.25747).
\end{abstract}

\maketitle
{\it Introduction---}The discrete-time quantum walk, a quantum counterpart to the classical random walk,
is a mathematical model of a particle that evolves in discrete-time steps via two unitary operators, coin-flipping and position-shifting~\cite{aharonov1993quantum, kempe2003quantum}.
In each step, the coin-flipping operator ($\mathcal{C}$) changes the {\it coin} state of the particle, which is spanned by $|0\rangle_\co$ and $|1\rangle_\co\in \mathbb{C}^2$.
The subsequent position-shifting operator ($\mathcal{S}$) moves the particle to another position based on the outcome of the coin-flipping.
Since the coin state may form a quantum superposition of both $|0\rangle_\co$ and $|1\rangle_\co$,
the particle simultaneously moves in different directions and becomes widely distributed across space as a quantum superposition.

Thanks to its unique quantum spatial distribution, the discrete-time quantum walk plays a crucial role in several quantum algorithms.
Representative examples are quantum walk-based search algorithms in computational space,
structured as hypercubes~\cite{shenvi2003quantum, potovcek2009optimized, hein2009quantum},
multi-dimensional lattices~\cite{ambainis2004coins, portugal2013quantum},
bipartite graphs~\cite{szegedy2004quantum, rhodes2019quantum, peng2024deterministic},
or more complicated graphs~\cite{berry2010quantum, bezerra2021quantum, mukai2020discrete}.
They provide a quadratic speed-up over classical counterparts when searching for the target data in the computational space.

Implementations of quantum computation are also important applications of the discrete-time quantum walk.
One example is a universal quantum computer based on a single-particle discrete-time quantum walk~\cite{lovett2010universal}.
The architecture consists of many wires and graphs through which a flying qubit propagates in discrete time steps.
Another example is an application to quantum random access memory ~\cite{giovannetti2008quantum}.
In discrete-time steps, the register spreads over multiple memory cells through quantum spatial distribution and retrieves data from them in parallel~\cite{asaka2021quantum,asaka2023two2}.

Here, this work incorporates neighboring-interaction operator ($\mathcal{N}$) into the discrete-time quantum walk
to account for many-body effects among multiple particles,
and then further extends the versatility of the quantum walk to quantum error correction,
yielding two advantages for the implementation of the error-correcting scheme.
The first benefit is resource efficiency;
multi-qubit redundant encoding of information can be achieved with a small number of particles.
Indeed, we demonstrate that our scheme enables the implementation of Shor's nine-qubit code using only three particles
with two ancillary particles used for the syndrome measurement.
The second benefit is implementation feasibility;
encoding qubit information and correcting errors can be implemented with interactions only between nearest-neighbor particles.

Furthermore, our proposal would lead to ultrafast quantum error correction,
significantly mitigating the risk of encoded information being exposed to noise.
Namely, the scheme mainly consists of iterations of $\mathcal{C}\rightarrow\mathcal{S}\rightarrow\mathcal{N}\rightarrow\mathcal{C}\rightarrow\cdots$,
which would be connected to continuous-time multi-particle dynamics by taking the continuous limit of these discrete-time steps.
Indeed, whereas the incorporation of many-body effects has yet to be achieved,
the researchers have succeeded in connecting the discrete-time quantum walk, i.e., iterations of $\mathcal{C}\rightarrow\mathcal{S}\rightarrow\mathcal{C}\rightarrow\cdots$,
to the continuous-time quantum walk~\cite{strauch2006connecting,childs2010relationship,d2010connection,mn2015continuous,manighalam2019continuum} and the Dirac equation~\cite{chandrashekar2010relationship,di2012discrete,debbasch2013discrete,d2016discrete,manighalam2019continuum} by taking the continuous limit of the iterations.

\indent
{\it Multi-particle quantum walk---}We begin by formulating the dynamics of $3+2$ particles on nested squares.
Here, each particle mainly evolves through the iterative application of $\mathcal{C}$, $\mathcal{S}$, and $\mathcal{N}$, all introduced in this section.
The purpose of this setting is to enable ultrafast error correction by taking the continuous limit of the steps,
whereas such reformulation is out of this paper's scope.

\begin{figure}[b]
    \centering
    \vspace{-0.5cm}
    \begin{minipage}[b]{0.49\columnwidth}
        \centering
        \includegraphics[height=2.9cm]{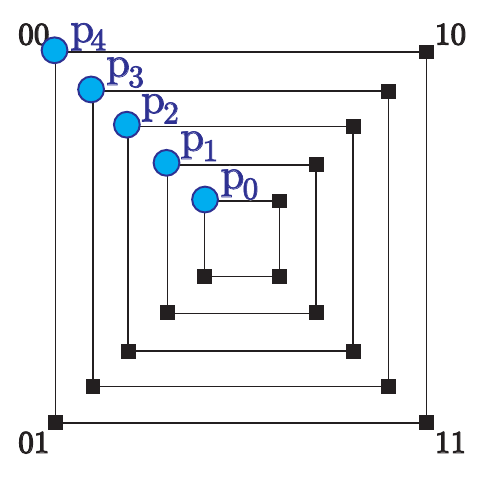}
        \\
        (a)
        \label{figure: }
    \end{minipage}
    \begin{minipage}[b]{0.49\columnwidth}
        \centering
        \includegraphics[height=2.9cm]{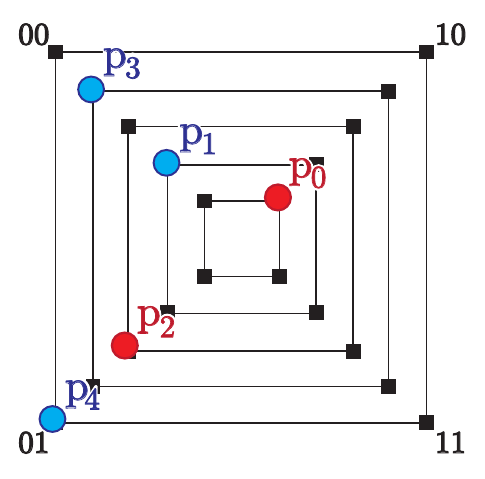}
        \\
        (b)
        \label{figure: }
    \end{minipage}
    \caption{The three-particle system $\{\pa{0}, \pa{2}, \pa{4}\}$ with ancillary particles $\{\pa{1},\pa{3}\}$ for syndrome measurement.
    The state in (a) is written as $\left(|0\rangle_\co |00\rangle_\po\right)_{\pa{4}} \left(|0\rangle_\co |00\rangle_\po\right)_{\pa{2}} \left(|0\rangle_\co |00\rangle_\po\right)_{\pa{0}}$,
    and in (b) as $\left(|0\rangle_\co |10\rangle_\po\right)_{\pa{4}} \left(|1\rangle_\co |01\rangle_\po\right)_{\pa{2}} \left(|1\rangle_\co |01\rangle_\po\right)_{\pa{0}}$,
    where blue represents the state $|0\rangle_\co$ while red represents $|1\rangle_\co$.
    }
    \label{figure: quantum walk on square lattices}
\end{figure}

In our model, which is an extension of the model in Ref.~\cite{singh2021universal,chawla2023multi},
all particles have both coin and position states, spanned respectively by $\{|0\rangle_\co, |1\rangle_\co\}\subset \mathbb{C}^2$
and $\{|00\rangle_\po, |10\rangle_\po, |11\rangle_\po, |01\rangle_\po\}\subset \mathbb{C}^4$ with the xy-coordinate rule.
We assume five particles/squares to demonstrate the implementation of Shor's nine-qubit code.
The particles are labeled $\pa{0}$--$\pa{4}$ from the innermost to the outermost square (Fig.~\ref{figure: quantum walk on square lattices}).
The set of particles $\{\pa{0},\pa{2}, \pa{4}\}$ will be used to host the redundantly encoded single-qubit information that is protected by our error-correcting scheme,
whereas $\{\pa{1},\pa{3}\}$ serves as an ancillary particles for error detection.

By the sequence $\mathcal{C}\rightarrow\mathcal{S}\rightarrow\mathcal{N}$,
every particle may or may not shift clockwise to the next vertex, and if a neighboring vertex is occupied by another particle, the two interact and acquire a phase of $-1$.
Specifically, each particle first updates its coin state through the coin-flipping operation ($\mathcal{C}$), depending on the vertex this particle occupies.
The subsequent position-shifting $(\mathcal{S}$) either keeps the particle stationary or moves it to the next vertex in a clockwise direction on the same square,
depending on whether the coin state is $|0\rangle_\co$ or $|1\rangle_\co$.
Finally, each pair of particles on adjacent squares acquires a phase factor of $-1$ through neighboring interactions ($\mathcal{N}$)
if they have the same coin state and occupy neighboring vertices, i.e., vertices on adjacent squares that share the same $x$ and $y$ coordinates.

Explicitly, all particles evolve under the action of the following three operators:
\begin{widetext}
    \begin{equation}
        \begin{array}{c}
            \displaystyle
            \mathcal{C}  := \prod_{i=0}^{4} \Biggl(\sum_{xy=00}^{11} U_\co^{(i;xy)} \otimes |xy\rangle \langle xy|_\po\Biggr)_{\pa{i}},\
            \mathcal{S} := \prod_{i=0}^{4} \Biggl(|0\rangle \langle 0|_\co \otimes I_\po + |1\rangle\langle 1|_\co \otimes R_\po \Biggr)_{\pa{i}},
            \\
            \displaystyle
            \mathcal{N} := \prod^3_{i=0} \Biggl((I_\co \otimes I_\po)^{\otimes 2} - 2 \sum_{c=0}^1 \sum_{xy=00}^{11} \Bigl(|c\rangle\langle c|_\co \otimes |xy\rangle\langle xy|_\po \Bigr)^{\otimes 2} \Biggr)_{\pa{i},\pa{i+1}}.
        \end{array}
        \label{eq: dynamics}
    \end{equation}
\end{widetext}
The right-side subscript ``$\pa{i}$'' (resp.~``$\pa{i}, \pa{i+1}$'') indicates that
the operator inside the parentheses acts nontrivially on the particle $\pa{i}$ (resp.~the particles $\pa{i}$ and $\pa{i+1}$).
The component $U_\co^{(i;xy)} \in \textrm{End}(\mathbb{C}^2)$ thus represents a unitary operator acting on the coin state of the corresponding particle.
In our error-correcting and encoding scheme, this takes the identity $I_\co := |0\rangle \langle 0|_\co + |1\rangle \langle 1|_\co$,
or the Pauli $X$ operator $X_{\co} :=  |1\rangle \langle 0|_\co + |0\rangle \langle 1|_\co$, depending on the square $i$ and vertex $xy$.
Additionally, the operator $I_\po \in \textrm{End}(\mathbb{C}^4)$ is the identity for the position state, meaning that it keeps the particle stationary.
The operator $R_\po$ is defined as $R_\po := |10\rangle \langle 00|_\po + |11\rangle \langle 10|_\po + |01\rangle \langle 11|_\po + |00\rangle \langle 01|_\po \in \textrm{End}(\mathbb{C}^4)$,
which moves the particle in a clockwise direction.

\indent
{\it Error model---}We introduce two types of unintended unitary operations acting on a particle, which arise naturally in our model: a coin-flipping error $E_\mathcal{C}$ and a position-shifting error $E_\mathcal{S}$.
They are defined as
\begin{align}
    E_\mathcal{C} =   & \sum_{x,y\in\{0,1\}} E_\co^{(xy)} \otimes |xy\rangle\langle xy|_\po,
    \label{eq: coin-flipping error}
    \\
    E_\mathcal{S}   = & \sum_{j\in\{0,1\}} |c\rangle\langle c|_\co \otimes \left(\alpha_j I_\po + \beta_j R_\po + \gamma_j R^\top_\po\right)
    \label{eq: position-shifting error}
\end{align}
where $E_\co^{(xy)}\in\textrm{End}(\mathbb{C}^2)$.
The coefficients $\alpha_j, \beta_j, \gamma_j \in \mathbb{C}$ ($j=0,1$) are chosen such that  $E_\mathcal{S}$ is unitary.
The error-correcting scheme in the subsequent section corrects either of these errors occurring in a particle.

Throughout this paper, we assume that these two types of errors do not occur simultaneously, as they originate from different physical sources.
Specifically, the error $E_{\mathcal{C}}$ is attributed to a malfunction in the coin-flipping operation $\mathcal{C}$ or disturbances from external noise.
In contrast, $E_{\mathcal{S}}$ can be regarded as an unintended position shift that either advances one step ahead or lags one step behind the intended position,
which may depend on the coin state or occur independently of it.

Here, the unified noise model formulated in the companion paper~\cite{asaka2026toward} is also correctable under the present error-correcting scheme,
as the stabilizer settings are identical to those in this paper.
This noise model contains not only $E_{\mathcal{C}}$ and $E_{\mathcal{S}}$, but also three types of noise sources inherent in each particle on the nested squares:
(i) arbitrary decoherence of the coin state, (ii) arbitrary decoherence of the position state,
and (iii) complete or partial loss, i.e., dephasing, of the superposition of spin and position states.

\indent
{\it Quantum walk error correction---}In what follows, we show that iterations of $\mathcal{C}$, $\mathcal{S}$, and $\mathcal{N}$ correct either the error $E_\mathcal{C}$ or $E_\mathcal{S}$ in at most one particle within the system $\{\pa{0}$, $\pa{2}$, $\pa{4}\}$.
Specifically, the error-correcting process mainly consists of three iterations of $(\mathcal{N}\mathcal{S}\mathcal{C})^6$.
Here, we measure the coin states of the measurement particles $\pa{1}$ and $\pa{3}$ after each iteration,
thereby sequentially collecting the eigenvalues of the stabilizer generators
$(s_0, s_2)$, $(s_1,s_3)$, and $(s_4,s_5)$ in Table~\ref{table: stabilizer generators, logical operators, and gauge transformations}.

Note that the states of $\{\pa{0}, \pa{2}, \pa{4}\}$ that the following scheme can protect against either $E_\mathcal{C}$~[Eq.~\eqref{eq: coin-flipping error}] or $E_\mathcal{S}$~[Eq.~\eqref{eq: position-shifting error}]
are limited to only the $s_0$--$s_5$ simultaneous eigenstates.
However, this is sufficient for the purpose of quantum error correction.
Namely, as described in the next section, we can redundantly encode single-qubit information into this system as the simultaneous eigenstate.

Measuring the $s_0$--$s_5$ eigenvalues, known as {\it syndrome measurement}, achieves twofold objectives.
First, the measurement projects $E_\mathcal{C}$ or $E_\mathcal{S}$ in the particle $\pa{0}$, $\pa{2}$, or $\pa{4}$ into bit- and phase-flip errors in the same particle.
Second, we identify the projected errors based on how the eigenvalues have changed, as shown in Table~\ref{table: correspondence between syndrome patterns and errors}.
The underlying mechanism is that both $E_\mathcal{C}$ and $E_\mathcal{S}$ are linear combinations of products of bit- and phase-flip errors.
Here, each product maps the state of $\{\pa{0}$, $\pa{2}$, $\pa{4}\}$ into a different $s_0$--$s_5$ simultaneous eigenspace, since the product anti-commutes with some of the generators $s_0$--$s_5$.

Note that an error that is part of the linear combination of $E_\mathcal{C}$ or $E_\mathcal{S}$ but not listed in Table~\ref{table: correspondence between syndrome patterns and errors} can be regarded as equivalent to one of the errors
in this table via the gauge transformations and stabilizer generators in Table~\ref{table: stabilizer generators, logical operators, and gauge transformations}.
For example, because $(I_\co Z_\textrm{x} I_\textrm{y})_{\pa{2}} = g_0^Z s_0 s_2 (Z_\co I_\textrm{x} I_\textrm{y})_{\pa{2}}$,
the error $(I_\co Z_\textrm{x} I_\textrm{y})_{\pa{2}}$ in $E_\mathcal{S}$ is treated as $(Z_\co I_\textrm{x} I_\textrm{y})_{\pa{2}}$ under the error correction.
Namely, both errors result in the same syndrome pattern and have an equivalent effect on the system $\{\pa{0},\ \pa{2},\ \pa{4} \}$ where single-qubit information is encoded.
We may refer to such equivalences as {\it gauge symmetry}, following the terminology in~\cite{poulin2005stabilizer},
and we will explain the reason for this gauge symmetry in the next section.

\begin{table}[t]
    \caption{Stabilizer generators $s_i\ (0\leq i\leq 5)$, gauge transformations $(g_i^Z, g_i^X)\ (0\leq i \leq 1)$,
        and the logical Z and X operators $(\bar{X},\bar{Y})$.
        The symbols $Z$ and $X$ represent the Pauli $Z$ and $X$ operators, respectively,
        and the subscripts $\co$, $\mathrm{x}$, or $\mathrm{y}$ indicate the state of the corresponding particle on which the operator acts.
    }
    \label{table: stabilizer generators, logical operators, and gauge transformations}
    \begin{tabular}{lccccccccccc}
        \hline
        $s_0  =$     & $(I_\co$ & $I_\textrm{x}$ & $I_\mathrm{y})_{\pa{4}}$ & $\otimes$ & $(Z_\co$ & $Z_\mathrm{x}$ & $I_\mathrm{y})_{\pa{2}}$ & $\otimes$ & $(Z_\co$ & $Z_\mathrm{x}$ & $I_\mathrm{y})_{\pa{0}}$ \\
        $s_1  =$     & $(I_\co$ & $I_\textrm{x}$ & $I_\mathrm{y})_{\pa{4}}$ & $\otimes$ & $(Z_\co$ & $I_\mathrm{x}$ & $Z_\mathrm{y})_{\pa{2}}$ & $\otimes$ & $(Z_\co$ & $I_\mathrm{x}$ & $Z_\mathrm{y})_{\pa{0}}$ \\
        $s_2  =$     & $(Z_\co$ & $Z_\textrm{x}$ & $I_\mathrm{y})_{\pa{4}}$ & $\otimes$ & $(Z_\co$ & $Z_\mathrm{x}$ & $I_\mathrm{y})_{\pa{2}}$ & $\otimes$ & $(I_\co$ & $I_\mathrm{x}$ & $I_\mathrm{y})_{\pa{0}}$ \\
        $s_3  =$     & $(Z_\co$ & $I_\textrm{x}$ & $Z_\mathrm{y})_{\pa{4}}$ & $\otimes$ & $(Z_\co$ & $I_\mathrm{x}$ & $Z_\mathrm{y})_{\pa{2}}$ & $\otimes$ & $(I_\co$ & $I_\mathrm{x}$ & $I_\mathrm{y})_{\pa{0}}$ \\
        $s_4  =$     & $(I_\co$ & $I_\textrm{x}$ & $I_\mathrm{y})_{\pa{4}}$ & $\otimes$ & $(X_\co$ & $X_\mathrm{x}$ & $X_\mathrm{y})_{\pa{2}}$ & $\otimes$ & $(X_\co$ & $X_\mathrm{x}$ & $X_\mathrm{y})_{\pa{0}}$ \\
        \vspace{0.33pt}
        $s_5  =$     & $(X_\co$ & $X_\textrm{x}$ & $X_\mathrm{y})_{\pa{4}}$ & $\otimes$ & $(X_\co$ & $X_\mathrm{x}$ & $X_\mathrm{y})_{\pa{2}}$ & $\otimes$ & $(I_\co$ & $I_\mathrm{x}$ & $I_\mathrm{y})_{\pa{0}}$ \\
        $g^Z_0  =$   & $(Z_\co$ & $Z_\textrm{x}$ & $I_\mathrm{y})_{\pa{4}}$ & $\otimes$ & $(Z_\co$ & $Z_\textrm{x}$ & $I_\mathrm{y})_{\pa{2}}$ & $\otimes$ & $(Z_\co$ & $Z_\textrm{x}$ & $I_\mathrm{y})_{\pa{0}}$ \\
        $g^Z_1  =$   & $(Z_\co$ & $I_\textrm{x}$ & $Z_\mathrm{y})_{\pa{4}}$ & $\otimes$ & $(Z_\co$ & $I_\textrm{x}$ & $Z_\mathrm{y})_{\pa{2}}$ & $\otimes$ & $(Z_\co$ & $I_\textrm{x}$ & $Z_\mathrm{y})_{\pa{0}}$ \\
        $g^X_0  =$   & $(X_\co$ & $I_\textrm{x}$ & $X_\mathrm{y})_{\pa{4}}$ & $\otimes$ & $(X_\co$ & $I_\mathrm{x}$ & $X_\mathrm{y})_{\pa{2}}$ & $\otimes$ & $(X_\co$ & $I_\mathrm{x}$ & $X_\mathrm{y})_{\pa{0}}$ \\
        \vspace{0.33pt}
        $g^X_1  =$   & $(X_\co$ & $X_\textrm{x}$ & $I_\mathrm{y})_{\pa{4}}$ & $\otimes$ & $(X_\co$ & $X_\mathrm{x}$ & $I_\mathrm{y})_{\pa{2}}$ & $\otimes$ & $(X_\co$ & $X_\mathrm{x}$ & $I_\mathrm{y})_{\pa{0}}$ \\
        $\bar{Z}  =$ & $(Z_\co$ & $Z_\textrm{x}$ & $Z_\mathrm{y})_{\pa{4}}$ & $\otimes$ & $(Z_\co$ & $Z_\mathrm{x}$ & $Z_\mathrm{y})_{\pa{2}}$ & $\otimes$ & $(Z_\co$ & $Z_\mathrm{x}$ & $Z_\mathrm{y})_{\pa{0}}$ \\
        $\bar{X}  =$ & $(X_\co$ & $X_\textrm{x}$ & $X_\mathrm{y})_{\pa{4}}$ & $\otimes$ & $(I_\co$ & $I_\mathrm{x}$ & $I_\mathrm{y})_{\pa{2}}$ & $\otimes$ & $(I_\co$ & $I_\mathrm{x}$ & $I_\mathrm{y})_{\pa{0}}$ \\
        \hline
    \end{tabular}
\end{table}

\begin{table}[t]
    \caption{The correspondence between the measured stabilizer syndrome and the operator mapped from a nontrivial unitary error on particle $\pa{0}$, $\pa{2}$, or $\pa{4}$ by the syndrome measurement.
        The symbol $m_i \in \{0, 1\}\ (0\leq i \leq 5)$ indicates that the measured eigenvalue of the stabilizer generator $s_i$ is $(-1)^{m_i}$.
    }
    \label{table: correspondence between syndrome patterns and errors}
    \centering
    \hspace{-40pt}
    \begin{minipage}{0.255\textwidth}
        \centering
        \begin{tabular}{c|c|c}
            \hline
            $m_5$ & $m_4$ & Phase flip                                     \\
            \hline
            0     & 0     & None                                           \\
            0     & 1     & $(Z_\co\ I_\textrm{x}\ I_\mathrm{y})_{\pa{0}}$ \\
            1     & 0     & $(Z_\co\ I_\textrm{x}\ I_\mathrm{y})_{\pa{4}}$ \\
            1     & 1     & $(Z_\co\ I_\textrm{x}\ I_\mathrm{y})_{\pa{2}}$
            \\
            \hline
        \end{tabular}
    \end{minipage}
    \begin{minipage}{0.18\textwidth}
        \centering
        \begin{tabular}{cccc|c}
            \hline
            $m_3$ & $m_2$ & $m_1$ & $m_0$ & Bit flip                                       \\
            \hline
            0     & 0     & 0     & 0     & None                                           \\
            0     & 0     & 0     & 1     & $(I_\co\ X_\textrm{x}\ I_\mathrm{y})_{\pa{0}}$ \\
            0     & 0     & 1     & 0     & $(I_\co\ I_\textrm{x}\ X_\mathrm{y})_{\pa{0}}$ \\
            0     & 0     & 1     & 1     & $(X_\co\ I_\textrm{x}\ I_\mathrm{y})_{\pa{0}}$
            \\
            \hline
        \end{tabular}
    \end{minipage}
    \\
    \begin{minipage}{0.23\textwidth}
        \centering
        \vspace{5pt}
        \begin{tabular}{cccc|c}
            \hline
            $m_3$ & $m_2$ & $m_1$ & $m_0$ & Bit flip                                       \\
            \hline
            0     & 1     & 0     & 0     & $(I_\co\ X_\textrm{x}\ I_\mathrm{y})_{\pa{4}}$ \\
            1     & 0     & 0     & 0     & $(I_\co\ I_\textrm{x}\ X_\mathrm{y})_{\pa{4}}$ \\
            1     & 1     & 0     & 0     & $(X_\co\ I_\textrm{x}\ I_\mathrm{y})_{\pa{4}}$
            \\
            \hline
        \end{tabular}
    \end{minipage}
    \hspace{5pt}
    \begin{minipage}{0.23\textwidth}
        \vspace{5pt}
        \centering
        \begin{tabular}{cccc|c}
            \hline
            $m_3$ & $m_2$ & $m_1$ & $m_0$ & Bit flip                                       \\
            \hline
            0     & 1     & 0     & 1     & $(I_\co\ X_\textrm{x}\ I_\mathrm{y})_{\pa{2}}$ \\
            1     & 0     & 1     & 0     & $(I_\co\ I_\textrm{x}\ X_\mathrm{y})_{\pa{2}}$ \\
            1     & 1     & 1     & 1     & $(X_\co\ I_\textrm{x}\ I_\mathrm{y})_{\pa{2}}$
            \\
            \hline
        \end{tabular}
    \end{minipage}
\end{table}

In the first step of the syndrome measurement, we determine the  $s_0$ and $s_2$ eigenvalues.
Initially, we allocate both ancillary particles $\pa{1}$ and $\pa{3}$ at the vertices labeled $00$.
Additionally, we set the components of $\mathcal{C}$ as $U_\co^{(i; xy)} = X_\co\ (i=1,3,\ xy=10,11$), with the others as identities.
Subsequently,
we repeat  $\mathcal{C}\rightarrow \mathcal{S}\rightarrow \mathcal{N}$ over six times,
with Hadamard operations applied to the coins at both the beginning and the end of the process
to record the $s_0$ and $s_2$ eigenvalues into the coin states of $\pa{1}$ and $\pa{3}$, respectively, as $|0\rangle_\co$ for $+1$ and $|1\rangle_\co$ for $-1$.
This procedure is written as
\begin{align}
     & (H_\co)_{\pa{1},\pa{3}}^{\otimes 2} \left(\mathcal{N}\mathcal{S}\mathcal{C}\right)^6 (H_\co)_{\pa{1},\pa{3}}^{\otimes 2},
    \label{eq: QW for stabilizer generators s_0 and s_2}
\end{align}
with the identities omitted.
Here, the subscripts of the tensor power ($(H_\co)^{\otimes 2}_{\pa{1},\pa{3}}$) each denote a target particle to which the operator is applied.
Finally, we obtain $m_0$ and $m_2$ ($m_i\in \{0, 1\}$ for $0\leq i \leq 5$) in Table~\ref{table: correspondence between syndrome patterns and errors} by measuring the coin states of $\{\pa{1}, \pa{3}\}$,
both returned to vertex $00$

The second step determines the $s_1$ and $s_3$ eigenvalues.
Here, the particles $\{\pa{0}, \pa{2}, \pa{4}\}$ are initially shifted by two from their original states just before the previous step for $s_0$ and $s_2$ begins.
For example, a particle originally at the vertex labeled $00$ with $|1\rangle_\co$ is now at the vertex labeled $11$.
From this condition, the procedure for collecting the eigenvalues follows the same flow as Eq.~\eqref{eq: QW for stabilizer generators s_0 and s_2},
but the components of $\mathcal{C}$ are set as $U_\co^{(i; xy)} = X_\co\ (i=1, 3,\ xy=11, 01)$.
Noe that the shifts by two of the particles $\pa{0}$, $\pa{2}$, and $\pa{4}$ are resolved through this step, and they have returned to their original positions.

The final step is determining the remaining $s_4$ and $s_5$ eigenvalues.
The procedure for this is given as
\begin{align}
     & (H_\co)_{\pa{1},\pa{3}}^{\otimes2}(H_\co H_\x H_\y)_{\pa{0},\pa{2},\pa{4}}^{\otimes 3}   \notag                                                            \\
     & \quad \mathcal{S}^2\left(\mathcal{N}\mathcal{S}\mathcal{C}\right)^6(H_\co H_\x H_\y)_{\pa{0},\pa{2},\pa{4}}^{\otimes 3}(H_\co)_{\pa{1},\pa{3}}^{\otimes2},
    \label{eq: QW for stabilizer generators s_4 and s_4}
\end{align}
where components of $\mathcal{C}$ are set as $U_\co^{(i; xy)} = X_\co\ (i=1, 3,\ xy=10, 01)$.
Here, the two applications of $\mathcal{S}$ after the six iterations are to resolve the aforementioned displacement of the physical particles, i.e., the shift by two.
Meanwhile, the gate operations $H_\co H_\x H_\y$ on the physical particles are to convert
$X_{\co}X_{\mathrm{x}}X_{\mathrm{y}}$ eigenstates of them into the $Z_{\co}Z_{\mathrm{x}}Z_{\mathrm{y}}$ eigenstates with the same eigenvalues, and vice versa,
where $H_\x$ and $H_\y$ denote the Hadamard gate acting on the $x$- and $y$-axis position state of a particle
(such Hadamard would be realized a quantum tunneling as Ref.~\cite{yamamoto2012electrical}).

In summary, through the three rounds of the six iterations with additional operations,
we collect the syndrome pattern $m_0$--$m_5$, which specifies how the $s_0$--$s_5$ eigenvalues change from the previous stabilizer measurement.
The target of the ultrafast implementation by taking the continuous-time limit are the iterations $(\mathcal{N}\mathcal{S}\mathcal{C})^6$ and $\mathcal{S}^2(\mathcal{N}\mathcal{S}\mathcal{C})^6$.
During the former (resp.~latter) iterations,
a physical particle with $|1\rangle_\co$ goes around the square one and a half times (resp.~two times).

Note that we do not need to correct emerging flipping errors after each cycle of syndrome measurement.
Namely, the historical record of syndrome patterns always provides a way to correct the accumulated flipping errors,
as these errors all commute with the syndrome measurements and the recovery operations (see Ref.~\cite{asaka2026toward} for the details).

\indent
{\it Information encoding---}We now describe that iterations of $\mathcal{C}$, $\mathcal{S}$, and $\mathcal{N}$ can also redundantly encode single-qubit information into the system $\{\pa{0}, \pa{2}, \pa{4}\}$.
Because the encoded state is a simultaneous eigenstate of $s_0$--$s_5$ in Table~\ref{table: stabilizer generators, logical operators, and gauge transformations},
the information gains resilience against the errors $E_\mathcal{C}$~[Eq.~\eqref{eq: coin-flipping error}] and $E_\mathcal{S}$~[Eq.~\eqref{eq: position-shifting error}] under our previously discussed error-correcting scheme.
Explicitly, we represent the encoded state as $\alpha|0\rangle_\Lo + \beta|1\rangle_\Lo \in (\mathbb{C}^2)^{\otimes 9}\ (\alpha, \beta \in \mathbb{C})$,
where $|0\rangle_\Lo$ and $|1\rangle_\Lo$ lie in the same simultaneous eigenspace of $s_0$--$s_5$ and thus yield the same syndrome pattern.

To understand the specific form of these two states $|0\rangle_\Lo$ and $|1\rangle_\Lo$, it is important to consider that their simultaneous eigenspace consists of three virtual qubits~\cite{poulin2005stabilizer}
(this space naturally has the structure of $(\mathbb{C}^2)^{\otimes 3}$, because each of $s_0$--$s_5$ divides the nine-qubit Hilbert space equally into its $+1$ and $-1$ eigenstates).
One is referred to as the virtual logical qubit, whose computational $Z$- and $X$-basis states are defined by the anticommuting pair of logical operators $(\bar{Z}, \bar{X})$ in Table~\ref{table: stabilizer generators, logical operators, and gauge transformations}.
The remaining two are referred to as virtual gauge qubits, each of which has a pair of gauge transformations $(g_i^X, g_i^Z)\ (i\in\{0,1\})$ in the same table defining its computational basis states.

Notably, the two states  $|0\rangle_\Lo$ and $|1\rangle_\Lo$ are distinguished by which computational $Z$-basis state the virtual logical qubit takes.
Namely, they are $+1$ and $-1$ eigenstates of the logical operator $\bar{Z}$.
These states are, of course, protected by our error-correcting scheme as they are also $s_0$--$s_5$ simultaneous eigenstates.

Meanwhile, two virtual gauge qubits do not carry useful information but reflect the presence of the gauge symmetry.
Namely, they partly absorb a flipping error--a summand in $E_\mathcal{C}$ or $E_{\mathcal{S}}$--that the syndrome measurement cannot detect,
and then manifest it as a detectable error in Table~\ref{table: correspondence between syndrome patterns and errors}.
For example, because the transformation $g_0^Z$ acts only on one of the virtual gauge qubits,
our error-correcting scheme can, as mentioned earlier, treat the error $(I_\co Z_\textrm{x} I_\textrm{y})_{\pa{2}} (=g_0^Z s_0 s_2 (Z_\co I_\textrm{x} I_\textrm{y})_{\pa{2}})$
unlisted in this Table as equivalent to $(Z_\co I_\textrm{x} I_\textrm{y})_{\pa{2}}$ with $g_0^Z$ seemingly absorbed.
Here, $g_0^Z$ does not change the syndrome pattern as it commutes all stabilizer generators.
See the companion paper~\cite{asaka2026toward} for details.

To encode information into the virtual logical qubit of the system $\{\pa{0}, \pa{2}, \pa{4}\}$,
we place an additional square to the left of the outermost square where the particle $\pa{4}$ exists:
\begin{align*}
    \begin{array}{c}
        \includegraphics[width=6.4cm]{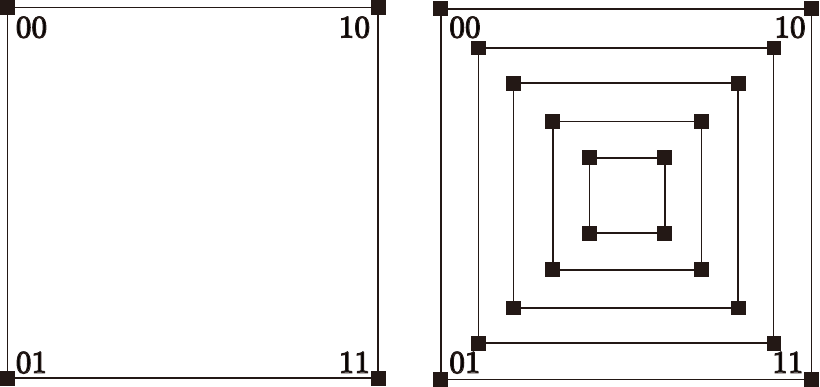}
    \end{array}
\end{align*}
Additionally, we set the external particle labeled $\pa{\ex}$ with coin state $|0\rangle_{\co}$ at the vertex $00$ of this extended square.
This external particle exhibits the same evolution as other particles that follow the operator $\mathcal{C}$, $\mathcal{S}$, and $\mathcal{N}$.
Here, the interaction between the adjacent pair of particles $\pa{\ex}$ and $\pa{4}$ occurs
when $\pa{\ex}$ is located at the vertex $10$ (resp.~$11$) and $\pa{4}$ at the vertex $00$ (resp.~$01$).

As a preliminary step for encoding, we initially prepare the logical state  $|0\rangle_{\Lo}$
by measuring the eigenvalues of the six stabilizer generators $s_0$--$s_5$ and the logical operator $\bar{Z}$.
From this point, the state $|0\rangle_{\Lo}$ is regarded as the obtained
simultaneous eigenstate corresponding to the measured syndrome and the eigenvalue of $\bar{Z}$
(if its eigenvalue of $\bar{Z}$ is reversed, the state becomes $|1\rangle_{\Lo}$).

Subsequently, by a trivial procedure, we prepare single-qubit information in the coin state of $\pa{\ex}$ located at the vertex $00$,
and then encode this information to the virtual logical qubit.
This encoding is specifically achieved in three stages:
\begin{align}
     & \bigl\{\left(\alpha |0\rangle_{\co} + \beta|1\rangle_{\co}\right) \bigr\}_{\pa{\ex}} |0\rangle_{\Lo}
    \label{eq: encoding (stage0)}
    \\
     & \xrightarrow{\textrm{(i)}}\alpha \left(|0\rangle_{\co}\right)_{\pa{\ex}}|0\rangle_{\Lo} + \beta \left(|1\rangle_{\co}\right)_{\pa{\ex}}|1\rangle_{\Lo}
    \label{eq: encoding (stage1)}
    \\
     & \xrightarrow{\textrm{(ii)}}  (|0\rangle_\co )_{\pa{\ex}} (a|0\rangle_\Lo + b |1\rangle_\Lo) / \sqrt{2} \notag
    \\
     & \qquad\qquad +(|1\rangle_\co  )_{\pa{\ex}} (a|0\rangle_\Lo - b |1\rangle_\Lo) / \sqrt{2}
    \label{eq: encoding (stage2)}
    \\
     & \xrightarrow{\textrm{(iii)}} (|c\rangle_\co )_{\pa{\ex}} (a|0\rangle_\Lo + b |1\rangle_\Lo)\quad (c\in\{0,1\}),
    \label{eq: encoding (stage3)}
\end{align}
where we abbreviate $(|c\rangle_\co |00\rangle_\po)_{\pa{\ex}}$ as $(|c\rangle_\co)_{\pa{\ex}} (c=0,1)$.
Here, (i) is applying the CNOT operation under which
the logical operator $\bar{X}$ flips the $\bar{Z}$ eigenvalue of the logical qubit
if and only if the coin state of $\pa{\ex}$ takes $|1\rangle_\co$:
\begin{align}
    \mathrm{CNOT} := \sum_{c=0}^1 (|c\rangle\langle c|_\co)_{\pa{\ex}} \otimes (\bar{X})^c.
    \label{eq: CNOT (control: coin, target: logical)}
\end{align}
The implementation of this gate is described in the next paragraph (see Eq.~\eqref{eq: CNOT (control: coin, target: logical)}).
Subsequently, (ii) is applying a Hadamard gate $H_\co$ to the particle $\pa{\ex}$,
and (iii) consists of measuring the coin state of $\pa{\ex}$ and applying the logical phase-flip $\bar{Z}$ to the system $\{\pa{0},\pa{2},\pa{4}\}$ if $|1\rangle_\co$ is measured.
Note that $\bar{Z}$ can be simply performed by applying the $Z_\co$ to these three particles,
given that $\bigotimes_{i\in \{0, 2, 4\}} \left(Z_{\co} I_{\mathrm{x}} I_{\mathrm{y}}\right)_{\pa{i}} = (s_0 s_1 g_0^Z g_1^Z) \bar{Z}$.

The implementation of the CNOT operation [(i) in Eq.~\eqref{eq: encoding (stage1)}] consists of the three types of quantum walk operators and Hadamard gates as
\begin{align}
    \mathrm{CNOT}= (H_\co H_\x H_\y)_\pa{4}  (\mathcal{N}\mathcal{S}\mathcal{C})^8  (H_\co H_\x H_\y)_\pa{4},
    \label{eq: CNOT implementation}
\end{align}
Here, the components of $\mathcal{C}$ are set as $U_\co^{(4;xy)}=X_\co$ ($xy=00,10,11,01$),
and under this, the eight iterations function as
\begin{align}
    (\mathcal{N}\mathcal{S}\mathcal{C})^8 = \sum_{c=0}^1 (|c\rangle\langle c|_\co)_\pa{\ex} \otimes (Z_\co Z_\x Z_\y)_\pa{4}.
    \label{eq: colo CP implementation}
\end{align}
During these iterations, a physical particle goes around the square up to twice.

\indent
{\it Conclusion---}We have proposed a novel quantum error-correcting scheme via the multi-particle discrete-time quantum walk.
Our scheme utilizes the quantum spatial distribution to implement the error-correcting code resource-efficiently and feasibly,
i.e., to implement Shor's nine-qubit code using only three particles while requiring only nearest-neighbor interactions.
Furthermore, the scheme mainly consists of iterations of three types of quantum walk operators as $\mathcal{C}\rightarrow\mathcal{S}\rightarrow\mathcal{N}\rightarrow\mathcal{C}\cdots$,
allowing us to take the continuous-time limit and to derive ultrafast quantum error correction.
The present study also offers the crucial insight that the gauge symmetry within the stabilizer formalism~\cite{gottesman1998theory,poulin2005stabilizer,dauphinais2024stabilizer}
is an essential ingredient for our scheme to achieve the resilience against dominant errors $E_\mathcal{C}$ and $E_\mathcal{S}$.

Here, the companion paper~\cite{asaka2026toward} further enhances this insight by providing a unified correctable noise model
which is otherwise uncorrectable without the symmetry.
Furthermore, this paper proposes an implementation of logical Toffoli and Hadamard gates using quantum spatial distribution over the nested squares;
if we succeed in reformulating the implementation as iterations of $\mathcal{C}\rightarrow\mathcal{S}\rightarrow\mathcal{N}\rightarrow\mathcal{C}\rightarrow\cdots$,
we would derive ultrafast, resource-efficient, and practically feasible fault-tolerant quantum computing.

\acknowledgements
We thank Kazumitsu Sakai for constructive discussions and his thoughtful advice on this paper.
This work was partially supported by JSPS KAKENHI Grant Numbers 23KJ1962 and 26K17055.

\bibliography{ref}
\end{document}